\documentclass[twocolumn,preprintnumbers,amsmath,amssymb,floatfix,superscriptaddress,nofootinbib]{revtex4}

\usepackage{amsmath,hyperref,amssymb}
\usepackage{graphicx}% Include figure files
\usepackage{dcolumn}% Align table columns on decimal point
\usepackage{bm}% bold math
\usepackage{verbatim}
\usepackage{tabularx}
\usepackage{slashed}
\usepackage{cancel}
\usepackage[boxsize=2em,aligntableaux=center]{ytableau}
\usepackage{float}

\def\be{\begin{equation}}
\def\ee{\end{equation}}
\def\bea{\begin{eqnarray}}
\def\eea{\end{eqnarray}}

\newcommand{\nn}{\nonumber}

\begin{document}

\vspace*{-30mm}

\title{Finding the Strong CP problem at the LHC}
\author{Raffaele Tito D'Agnolo}
\affiliation{School of Natural Sciences, Institute for Advanced Study, Princeton, New Jersey 08540, USA}
\author{Anson Hook}
\affiliation{School of Natural Sciences, Institute for Advanced Study, Princeton, New Jersey 08540, USA}

\begin{abstract}
We show that a class of parity based solutions to the strong CP problem predicts new colored particles with mass at the TeV scale, due to constraints from Planck suppressed operators.  The new particles are copies of the Standard Model quarks and leptons.  The new quarks can be produced at the LHC and are either collider stable or decay into Standard Model quarks through a Higgs, a W or a Z boson. We discuss some simple but generic predictions of the models for the LHC and find signatures not related to the traditional solutions of the hierarchy problem.  We thus provide alternative motivation for new physics searches at the weak scale. We also briefly discuss the cosmological history of these models and how to obtain successful baryogenesis.
 %Two simple but generic predictions of the models are that the new quarks are not singly produced at the LHC and that at large masses they decay as vector-like singlets, with a branching ratio to W bosons two times larger than that to Z or Higgs bosons.  We also briefly discuss the cosmological history of these models.

\end{abstract}

\vspace*{1cm}

\maketitle

%%%%%%%%%%%%%%%%%%%%%%%%%%%%%%%%%%%%%%%%%%
\section{Introduction}

The Standard Model (SM) of particle physics provides an excellent description of all known low energy phenomena.  However, there are several instances in the SM where our effective field theory intuition fails spectacularly. These are the cosmological constant, the Higgs mass (the hierarchy problem), the neutron electric dipole moment (the strong CP problem) and the Yukawa couplings.  These problems have motivated most of the work on extensions of the SM that are currently being probed experimentally.  The hierarchy problem has been the main driving force behind searches for new physics at the Large Hadron Collider (LHC).  The reason is that any dynamical explanation of the smallness of the Higgs mass requires TeV scale physics while the other problems do not.  In this paper, we note that certain solutions to the strong CP problem also provide strong motivation for new physics at the LHC.

The neutron electric dipole moment is proportional to 
\bea
\overline \theta = \theta + \text{arg} \, \text{det} \, Y_u Y_d
\eea
where $\theta$ is the coefficient of the CP violating term in the QCD action $G_{\mu\nu}^a\tilde G^{\mu\nu}_a$ and $Y_{u,d}$ are the Yukawa matrices.  Current experimental measurements of the neutron electric dipole moment indicate that $\overline \theta < 10^{-10}$~\cite{Baker:2006ts}, with an order of magnitude uncertainty from theory~\cite{Vicari:2008jw,Engel:2013lsa}.  This result is especially surprising given that the Yukawa matrices are complex and have an order one CKM phase, i.e. CP is badly broken in the SM.  The smallness of $\overline \theta$ is called the strong CP problem.

There are two broad categories of solutions to the strong CP problem.  The first are solutions based on anomalous symmetries.  The most well known of these solutions are the axion~\cite{Peccei:1977hh,Peccei:1977ur,Weinberg:1977ma,Wilczek:1977pj} and the massless up quark~\cite{'tHooft:1976up}.  In the UV, these solutions have an anomalous symmetry under which $\overline \theta$ shifts, rendering it unphysical.  In the IR, this anomalous symmetry is spontaneously broken and a scalar field dynamically removes $\overline \theta$ from the Lagrangian.

The second class of solutions are those which use Parity (P) or Charge-Parity (CP) to set $\overline \theta$ to zero in the UV. After P or CP is spontaneously broken, care must be taken to reintroduce a large CKM phase but a small $\overline \theta$.  The most well known of the CP based solutions are the Nelson-Barr approach~\cite{Nelson:1983zb,Barr:1984qx} and~\cite{Babu:1989rb}.  More recently, a systematic approach to the mediation of CP violation to the SM was done in Ref.~\cite{Vecchi:2014hpa}.  The focus of this paper will be on the parity based solutions~\cite{Barr:1991qx}.

The fact that the Strong CP problem can provide motivation for new physics at the TeV scale was first observed in Ref.~\cite{Hook:2014cda}.  There it was shown that in the context of a massless quark solution to the strong CP problem, higher dimensional operators combined with the stringent bounds on the neutron EDM can require the existence of new colored particles at the TeV scale.  In this note, we show that a broad class of parity based solutions to the strong CP problems are also subject to strong constraints from higher dimensional operators and also predict colored TeV scale physics.

%The class of parity based solutions to the strong CP problem that we study was first presented in Ref.~\cite{Barr:1991qx}.  In this paper we focus on the fact that higher dimensional operators make the theory testable at the LHC.

%%%%%%%%%%%%%%%%%%%%%%%%%%%%%%%%%%%%%%%%%%
\section{Model}

To solve the strong CP problem, we define a generalized parity and then spontaneously break it without introducing new phases.
We double the matter content of the SM and enlarge the gauge group to $SU(3)_c \times SU(2)_W \times SU(2)'_W \times U(1)_Y$.  Under generalized parity, $SU(3)_c \times U(1)_Y$ are invariant while $SU(2)_W$ and $SU(2)'_W$ are exchanged. The matter content of the SM is doubled so that generalized parity sends SM fermions to their mirror conjugates and our Higgs into the mirror Higgs. In what follows we denote the new particles with a prime. 
The $\theta$ angle is odd under generalized parity. This forces $\theta = 0$ for $SU(3)_c$ and $U(1)_Y$ in the UV. Furthermore, the symmetry results in the Yukawa matrices taking the form
\bea
\mathcal{L} &\supset& -Y_u H Q u^c - Y'_u H' Q' u'^c \nn \\
&=& -Y_u H Q u^c - Y_u^\star H' Q' u'^c, 
\eea
which gives
\bea
\overline \theta  = \text{arg} \, \text{det} \, Y_u Y_d + \text{arg} \, \text{det} \, Y'_u Y'_d = 0 \, .
\eea
Thus an exact generalized parity solves the strong CP problem while allowing a non-zero CKM phase at tree level.

At this point, generalized parity is a good symmetry of the theory.  However we do not see mirror quarks at low energies and this symmetry must be spontaneously broken.  To implement the spontaneous breaking, we assume that there exists a SM singlet pseudo-scalar $\phi$ which is odd under generalized parity and obtains a vacuum expectation value (vev).  The most general Lagrangian for the scalar $\phi$ is 
\bea
\mathcal{L}_\phi &=& m_\phi^2 \phi^2 - \lambda_\phi \phi^4 + \Lambda \phi ( H' H'^\dagger - H H^\dagger ) \nn \\
&-& \lambda \phi^2 (H H^\dagger + H' H'^\dagger)\, .
\eea
After $\phi$ obtains a vev, the two Higgs vevs are split and $\overline \theta$ will be regenerated through loops. However, as shown in Ref.~\cite{Ellis:1978hq}, this effect is unobservably small in the SM.  Similar considerations show that it is also negligible for these types of theories~\cite{Barr:1991qx}.  

The breaking of the symmetry introduces a new scale $f$ in the theory:
\bea
\langle \phi \rangle \approx \langle H' \rangle \equiv \frac{f}{\sqrt{2}} \gg \langle H \rangle \equiv \frac{v}{\sqrt{2}}\, .
\eea
In an attempt to only solve one problem at a time, we ignore the hierarchy problem $f\gg v$ in the following discussion.  
This hierarchy can be made radiatively stable by introducing supersymmetry into the model. Alternatively, we could take the approach of split SUSY~\cite{ArkaniHamed:2004fb} and assume that the anthropic principle solves the hierarchy problem while supersymmetry stabilizes $f$ relatively to $M_{\rm Pl}$. In this framework we would have traded the tuning of the neutron electric dipole moment for a tuning that can be be explained using anthropic arguments.

As it is well known in the case of axion models~\cite{Kamionkowski:1992mf, Kallosh:1995hi, Holman:1992us}, solutions to the strong CP problem are constrained by higher dimensional operators~\cite{Berezhiani:1992pq}. If we include the effects of gravity, our theory contains dimension five interactions in the form
\bea
&&\frac{g_s^2\phi}{32 \pi^2 M_{\rm Pl}}G_{\mu\nu}^a\tilde G^{\mu\nu}_a\, , \, (Y_u H Q u^c + Y_u^\star H' Q' u'^c) \frac{i \phi}{M_{\rm Pl}}\, , \, \cdots \nn
\eea
After $\phi$ obtains a vev, these operators (which are related by an anomalous field redefinition) result in a non-zero $\overline \theta$.  Requiring these new contributions to be smaller than the experimental bound gives $\langle \phi \rangle \approx \langle H' \rangle \lesssim 10^{-10} M_{\rm Pl} \approx 10^6$~TeV. Therefore the $u, d$ and electron partners have masses roughly below 10 TeV.  As mentioned before, due to theory uncertainties the bound on $\overline \theta$ is only valid to an order of magnitude, so that the upper bound on the masses of these new particles can vary between 1 and 100~TeV. This constraint does not depend on the details of the model discussed at the beginning of this section and applies to all solutions which double the SM matter content.  Thus much of the parameter space of these models is within reach of LHC searches for new colored particles. 

The only way to relax the upper bound is to introduce additional structure in the symmetry breaking sector. We could imagine that it contains two scalars $\phi_1$ and $\phi_2$ and a new hidden symmetry under which both scalars are charged, but $\phi_1 \phi_2$ is neutral. If only one of the two scalars is odd under generalized parity and both get a vev of $\mathcal{O}(f)$, we can repeat most of the discussion above replacing $\phi$ with $\phi_1 \phi_2$. In this new theory $f \lesssim 10^{-5} M_{\rm Pl}$. So the presence of new TeV scale colored fermions in this class of theories is not required but is still a general possibility worthy of attention.

%%%%%%%%%%%%%%%%%%%%%%%%%%%%%%%%%%%%%%%%%%
\section{Phenomenology}

The behavior of the mirror particles at colliders is largely determined by the tree-level mass mixing between the SM and the mirror sector
\bea
\label{Eq: mixing}
\mathcal{L} \supset -\mu_u u^c u'^c - \mu_d d^c d'^c - \mu_e e^c e'^c +h.c.
\eea
Invariance under generalized parity requires the $\mu$ matrices to be Hermitian. In the limit where these mass mixings go to zero, there is an enhanced symmetry (mirror baryon and lepton number).  It is thus technically natural for these mass terms to be small.  

If $\mu$ is non-zero, then the mirror quarks can decay.  In order to determine how the mirror quarks decay, we first perform a spurious SM flavor rotation to make the Yukawa matrices diagonal and real.  All of the flavor violation is now in the CKM matrix and $\mu$. The mass matrix for the quarks can be easily diagonalized under the assumption that $\mu_{u,d}, y_{u,d} v \ll y_{u,d} f$. For example, in the case of up quarks we have 
\bea
\left(\begin{array}{c}u^c_m\\ u'^c_m \end{array}\right) &=& \left(\begin{array}{cc} 1-\frac{\epsilon_R \epsilon_R^\dagger}{2} & -\epsilon_R \\ \epsilon^\dagger_R &  1- \frac{\epsilon_R^\dagger \epsilon_R}{2} \end{array}\right) \left(\begin{array}{c}u^c\\ u'^c \end{array}\right) \nn \\
 \left(\begin{array}{c}u_m\\ u'_m \end{array}\right) &=& \left(\begin{array}{cc} 1 & -\epsilon_L^\star \\ \epsilon_L^T & 1 \end{array}\right) \left(\begin{array}{c}u\\ u' \end{array}\right) \nn \\
 \epsilon_L &\approx& \sqrt{2} \frac{v}{f^2} y_u \mu_u^\dagger y_u^{-1} y_u^{-1} \quad \epsilon_R \approx \sqrt{2} \frac{\mu_u^\dagger y_u^{-1}}{f}\, ,
\eea
where the subscript $m$ denotes mass eigenstates. The mixing of the left-handed SM quarks is suppressed compared to the right-handed mixing. This is not surprising since only the right-handed SM quarks mix directly with the new sector. The size of the suppression strongly depends on the generation indexes $\epsilon_L/\epsilon_R \approx 10^{-1} -10^{-7}$. 

The leading effect of $\epsilon_R$ is to shift Higgs couplings 
\bea
\mathcal{L} &\supset& -u y_u H u^c - d y_d H d^c = -u_m y_u H \left(1-\frac{\epsilon^u_R \epsilon_R^{u,\dagger}}{2}\right) u^c_m \nn \\
&-& u_m y_u H \epsilon_R u'^c_m + h.c. + \left(u\to d\right) +\mathcal{O}\left(\frac{1}{f^3}\right)\, , \label{eq:HigC}
\eea
while $Z$ boson couplings remain diagonal at tree-level and the $W$ boson is only affected by the smaller $\epsilon_L$ mixing. We find that the new contributions to Flavor Changing Neutral Currents (FCNC) in this model are dominated by tree-level Higgs exchange. The corresponding constraints on Higgs couplings have been worked out in Ref.~\cite{Blankenburg:2012ex}. Here we discuss only the bounds on the quark sector that are those relevant for LHC phenomenology and take $f\approx 10^8$~GeV, so that they apply to light quarks with $m\approx {\rm TeV}$. The constraints on the model are summarized in Tab.~\ref{tab:results}. 

If $\epsilon_R$ is a random matrix, we find that in order to avoid FCNC constraints, we need all its elements to satisfy $\epsilon_R^{u(d)} \lesssim 5(4)\times10^{-2}$. This is due to the fact that observables are not affected by a single element of the matrix, but by the sum $\sum_k (\epsilon_R)_{ik}(\epsilon_R^\dagger)_{kj}$. The assumption of a flavor anarchic $\epsilon_R$ requires that $\mu$ is a random matrix times the Yukawa matrices. This would be technically natural given the flavor symmetries, but it would introduce very diverse mass scales in the matrix $\mu$. However it is equally plausible that we have a single scale $\hat \mu$ and random $\mathcal{O}(1)$ flavor breaking parameters. In this case the bounds give $\hat \mu_{u(d)} \lesssim 190(100)\; {\rm GeV}$.
\begin{table}[!t]
\centering
\begin{tabular}{| c | c | c |}
\hline\hline
Operator & Observable & Bound \\ [0.5ex] 
\hline
$(\overline s_R d_L)^2$ & $\Delta m_K, \; \epsilon_K$ &  $\left(\epsilon_R^d \epsilon_R^{d,\dagger}\right)_{12} < 4.6\times10^{-3}$ \\
\hline
$(\overline c_R u_L)^2$ & $\Delta m_D, \;\phi_D,\; |q/p|$ &  $\left(\epsilon_R^u \epsilon_R^{u,\dagger}\right)_{12} < 6.1\times10^{-3}$ \\
\hline
$(\overline b_R d_L)^2$ & $\Delta m_{B_d},\; S_{B_d\to \psi K}$ &  $\left(\epsilon_R^d \epsilon_R^{d,\dagger}\right)_{13} < 6.4\times10^{-3}$ \\
\hline
$(\overline b_R s_L)^2$ & $\Delta m_{B_s}$ &  $\left(\epsilon_R^d \epsilon_R^{d,\dagger}\right)_{23} < 5.2\times10^{-2}$ \\
\hline
\end{tabular}
\caption{Various flavor constraints on the FCNC mixing parameter $\epsilon_R$ for $f=10^8$~GeV.  The constraints are taken from Ref.~\cite{Blankenburg:2012ex}.}
\label{tab:results}
\end{table}
As discussed above, aside from FCNC, the mixing between the two sectors allows for the mirror quarks to decay through the emission of a W, Z or Higgs boson. The decays are dominated by the couplings in the second line of Eq.~(\ref{eq:HigC}). There are three scenarios that are consistent with the flavor constraints that we have just discussed. The first consists in taking $\mu = 0$. In this case we have a new conserved mirror baryon and lepton number and massive stable charged particles at the LHC. In the second scenario we consider two inequivalent possibilities from the flavor perspective that have the same collider phenomenology, either $\epsilon_R$ is flavor anarchic, with $\epsilon_R \lesssim 4\times 10^{-2}$ or $\mu$ contains a single scale smaller than about $100$~GeV with random $\mathcal{O}(1)$ flavor violation. The mirror quarks then preferentially decay into third generation quarks. This is true also for more general choices of $\mu$ matrix elements as long as $\mu_{31}^{u(d)} > 1/y_{t(b)} (y_{c(s)} \mu_{21}^{u(d)}, y_{u(d)} \mu_{11}^{u(d)})$. In what follows we indicate this scenario as flavor anarchic $\epsilon_R$ or $\mu$. The third possibility is that $\mu$ is flavor diagonal and the first generation mirror quarks decay preferentially into first generation quarks. Similar signatures arise also if we take $\mu_{31}^{u(d)} < 1/y_{t(b)}(y_{c(s)} \mu_{21}^{u(d)}, y_{u(d)} \mu_{11}^{u(d)})$, with the possibility of having decays to second generation quarks. 

Before discussing the three cases in more detail, it is worth mentioning that the single production of first generation mirror quarks is suppressed by powers of a small Yukawa, making these processes unobservable at the LHC even for $\epsilon_R=\mathcal{O}(1)$. This is a generic feature of these models due to the doubling of $SU(2)_W$, which forces the mixing to proceed only in the right-handed SM sector. Therefore we show only bounds on final states arising from pair production of the mirror quarks.

\paragraph{$\mu = 0$}  In the first case, the mirror quarks are collider stable.  Currently the strongest bound is set by the ATLAS search in Ref.~\cite{ATLAS:2014fka}. To obtain a good estimate of the constraint on a $u'$ and a $d'$, we can use the stop and sbottom cross section exclusions. The ATLAS collaboration shows the results of two separate analyses, one that does not use the information from the muon system and is thus insensitive to the behavior of R-hadrons inside the calorimeters and one that exploits the data from the full detector. We use the two bounds and the heavy quark pair production cross section computed in~\cite{Chatrchyan:2013uxa} at NLO using {\tt HATHOR}~\cite{Aliev:2010zk} to get a mass exclusion for the $u'$ and $d'$.  We find $m_{u'} \gtrsim 1120$ GeV and $m_{d'} \gtrsim 1079$ GeV from the search not including the muon system. This is the most conservative bound and differs from the full detector exclusion by less than 20 GeV in both cases. The results from the CMS collaboration~\cite{Chatrchyan:2013oca} are similar. The cross section bounds set by CMS on stops are truncated at 1 TeV. However for $m\gtrsim 500$~GeV they asymptote to the gluino bounds. This is coincidental, but the trend is expected to continue for masses above 1 TeV, since the offline selection is fully efficient and the online selection has the same effect in the two cases\footnote{Private communication with Loic Quertenmont.}. Using the gluino bound and the heavy quark pair production cross section, we get $m_{u'} \gtrsim 1020$~GeV. Again this is the most conservative bound, coming from a tracker only search. Considering different hadronization models~\cite{Mackeprang:2009ad,Kraan:2004tz, Mackeprang:2006gx} and including the full detector can increase the exclusion by approximately 80 GeV.

\paragraph{$\epsilon_R$/$\mu$ flavor anarchic}

\begin{figure}[!t]
\begin{center}
\includegraphics[width=0.45\textwidth]{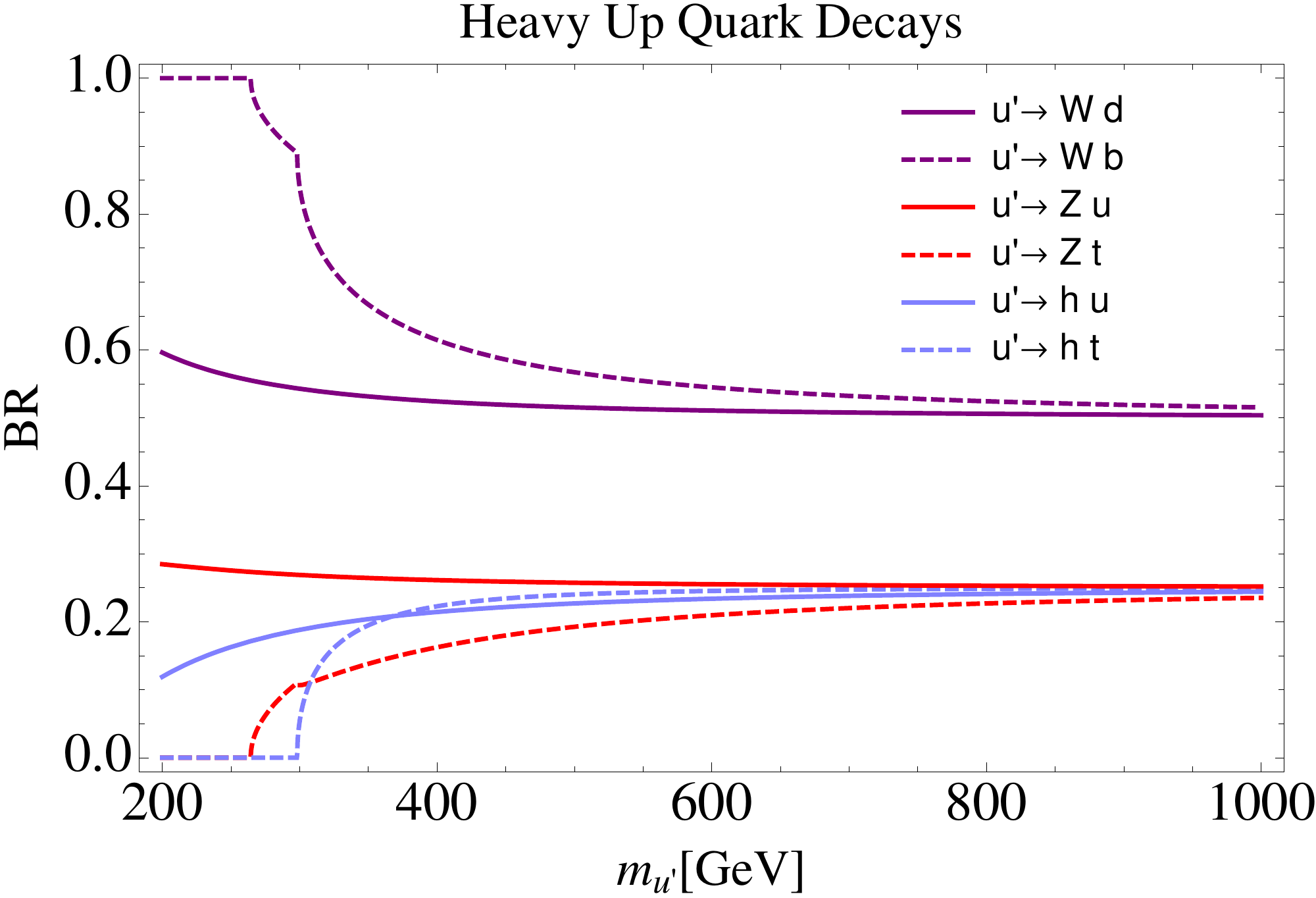}
\caption{$u'$ branching ratios as a function of mass for two different scenarios; the dashed (solid) lines show the $u'$ branching ratios when it decays only into third (first) generation quarks.  The purple, red and light blue lines represent the branching ratios for $u' \rightarrow W + b(d)$, $u' \rightarrow Z + t(u)$ and $u' \rightarrow h + t(u)$ respectively. In the large mass limit, they reduce to the values predicted by the Goldstone boson equivalence theorem.}
\label{Fig: BR}
\end{center}
\end{figure}

If $\epsilon_R$ is flavor anarchic the FCNC constraints discussed before apply and we require $\epsilon_R \lesssim 4\times 10^{-2}$. In the limit of large $m_{u'}$, we find the decay widths to be 
\bea
\label{Eq: decay width}
\Gamma (u' \rightarrow h + u_i) &\approx& \Gamma (u' \rightarrow Z + u_i)\approx \frac{1}{2} \Gamma (u' \rightarrow W^{\pm} + d_i) \nn \\
&\approx& y_{u_i}^2 (\epsilon_{R, u_i u'}^u)^2 \frac{m_{u'}}{32 \pi}\, ,
\eea
where we have shown the flavor indexes of the $\epsilon_R$ matrix.  The relation between the decay into the Higgs boson and the gauge bosons is given by the Goldstone boson equivalence theorem. 
If $\epsilon_R$ or $\mu$ are flavor anarchic, the size of the third generation Yukawas implies that the mirror up and down quarks preferentially decay into third generation SM quarks. The decays are prompt for $\epsilon_R \gtrsim 10^{-8}$.
These new heavy fermions look like top and bottom partners that are typically expected from Little Higgs and Composite Higgs type models. Both ATLAS and CMS have dedicated searches for these particles and their bounds are between $700$ and $800$~GeV. For these values of the heavy quark mass, the branching ratios respect the 2:1:1 relation from the Goldstone boson equivalence theorem to better than $10\%$ as shown in Fig.~\ref{Fig: BR}. The most stringent limit for a $u'$ decaying to third generation quarks is $m_{u'}\gtrsim 810$~GeV, set by the ATLAS leptons plus jets search in~\cite{ATLAS-CONF-2015-012}. For a down type heavy quark the bound is $m_{d'}\gtrsim 730$~GeV. In this case a CMS multilepton search~\cite{CMS:yut} and the same ATLAS search discussed above have comparable sensitivity.

\paragraph{$\mu$ approximately flavor diagonal}  If $\mu$ is flavor diagonal, the mirror quarks within LHC reach can only decay into first generation quarks. Fig.~\ref{Fig: BR} shows the branching ratios of a mirror up quark as a function of its mass. The decays are prompt for $\epsilon_R \gtrsim 10^{-3}$. The most recent direct search at the LHC was performed by CMS using $q' \to Wq$ final states~\cite{CMS:2014dka}. For the branching ratios in Fig.~\ref{Fig: BR} (${\rm BR}\left[q'\to Wq\right]\approx 0.5$) the analysis gives a bound weaker than $500$~GeV, the last mass point in the collaboration exclusion. A more systematic approach that includes signal regions sensitive to decays to $Z$ and Higgs bosons was adopted in~\cite{VanOnsem:2014dba}. Here $m_{u', d'} \gtrsim 530$~GeV is excluded for our values of the branching ratios, when the single production of the heavy quarks is subdominant. As discussed above, searches for singly produced mirror quarks~\cite{ATLAS:2012apa, Aad:2011yn, CMS:2013fea} are not sensitive to these models, while searches for three jet resonances~\cite{TheATLAScollaboration:2013xia}, multileptons~\cite{Chatrchyan:2014aea} or leptoquarks~\cite{CMS:zva} if recasted would give weaker bounds than those already shown~\cite{Blum:2015rpa, Delaunay:2013pwa}.

%%%%%%%%%%%%%%%%%%%%%%%%%%%%%%%%%%%%%%%%%%
\section{Cosmology}

A generic problem that all of these models face is the presence of domain walls associated with the spontaneous breaking of party.  Thus the reheating temperature of the universe should be below $f\approx10^8-10^9$ GeV.  
If the mirror particles are stable, their relic abundance overcloses the universe and the reheating temperature needs to be below $1-10$ TeV so that the mirror sector is not reheated. For the rest of the discussion we consider a mirror sector within LHC reach, i.e. $f\approx10^8$~GeV. If the mirror quarks are unstable, then they all decay before BBN for $\epsilon_R \gtrsim 10^{-13}(10^{-8})$ in the flavor anarchic (diagonal) case, a comparable constraint can be easily derived for mirror leptons.  The only potential problem are the mirror neutrinos.  
If we allow for an explicit breaking of lepton number, we can write the Lagrangian
\bea \label{Eq: neutrinos}
\mathcal{L}&\supset& -y_\nu \left ( H L \nu^c + H' L' \nu'^c \right ) - \frac{m_\nu}{2} \left (\nu^c \nu^c + \nu'^c \nu'^c \right ) - \nn \\
&-&\mu_\nu \nu^c \nu'^c +h.c.
\eea  
We can then integrate out the right-handed neutrinos and give mass to the light ones through the seesaw mechanism,  
\bea
\mathcal{L}&\supset& y_\nu^2 \frac{(HL) (HL)}{2 m_\nu} + y_\nu^2 \frac{(H'L') (H'L')}{2 m_\nu} \nn \\
&-& y_\nu^2 \frac{\mu_\nu (H L) (H'L')}{m_\nu^2}+h.c. +\mathcal{O}(1/m_\nu^3)\, .
\eea
In this setting, the mirror neutrinos freeze-out when relativistic and can overclose the universe. However they mix with the SM left-handed neutrinos and are heavy enough to decay into SM particles. Requiring the decays to take place before BBN gives $\mu_\nu \gtrsim 10^8$~GeV, where here and in the following we assume SM neutrino masses around one eV. Alternatively, lepton number could be preserved so that the Majorana mass terms in Eq.~(\ref{Eq: neutrinos}) are absent.  Integrating out $\nu^c$ and $\nu'^c$ gives the higher dimensional operator
\bea
\mathcal{L} \supset y_\nu^2\frac{(H' L') (H L)}{\mu_\nu}\, .
\eea
These mirror neutrinos act like right-handed neutrinos with a small dirac mass and freeze out when they are still relativistic. Their contribution to the effective number of relativistic degrees of freedom during BBN (recombination) is $\Delta N_\text{eff} = 0.14 (0.03)$, well below current constraints~\cite{Cooke:2013cba, Ade:2015xua}.

The $\mu \ne 0$ case has another rather remarkable property.  Since the mirror Higgs vev is so large, the quartic has run to a much smaller value when the electroweak phase transition takes place in the mirror sector and it is first order!  This allows for electroweak baryogenesis to proceed in the mirror sector.  Since the electroweak phase transition is already first order, we only need to introduce new CP violation such that the phase of the quark masses depends on the Higgs vev. This has the potential to ruin the solution to the strong CP problem.  For example, one could introduce CP violation of the form
\bea
H Q u^c \frac{i H^2}{\Lambda^2} - H' Q' u'^c \frac{i H'^2}{\Lambda^2}\, ,
\eea
with $\Lambda$ an order of magnitude or so above $f$ so that the CP violation is large enough to generate the observed baryon number asymmetry.  In this case after $H$ and $H'$ obtain different vevs, the phases of the mass matrices are no longer exactly opposite to each other and the Strong CP problem is reintroduced.  However, if the new particles which generate the Higgs dependent quark mass phases are parity even, we have quark masses that behave as
\bea
H Q u^c \frac{i (H^2+H'^2)}{\Lambda^2} - H' Q' u'^c \frac{i (H^2+H'^2)}{\Lambda^2} \, .
\eea
Even when $H$ and $H'$ get different vevs, the phases remain opposite to each other and the solution to the strong CP problem is unperturbed.

After successful electroweak baryogenesis in the mirror sector, the generated B+L asymmetry can be washed out by the sphalerons in our sector.  This happens if the baryon and lepton numbers in the two sectors are in chemical equilibrium or if the mirror sector particles decay before sphalerons freeze-out.  Both of these circumstances can be avoided by an appropriate choice of $\epsilon_R$.  We find that if the decays of the quarks take place when the SM temperature is between 100 GeV and an MeV, the mirror sector baryon number is naturally not in chemical equilibrium with the SM when electroweak sphalerons are active and the synthesis of light elements proceeds undisturbed.  If $\epsilon_R$ satisfies the requirement $10^{-25} \lesssim \sum_{q} y_{q}^2 (\epsilon_{R, q q'})^2 \lesssim 10^{-16}$, where $q'=u'$ or $d'$, then the mirror quarks decay in the right epoch. This choice automatically ensures that interactions of the type $q' q \to qq$ are not in equilibrium at high temperatures before the SM electroweak sphalerons freeze-out. For these values of the decay widths, we have decay lengths between a cm and $10^{10}$ cm. In a fraction of the parameter space where we can have successful baryogenesis, the mirror quarks appear as displaced vertices at the LHC~\cite{CMS:2014hka, CMS:2014wda, Aad:2015rba}. However this is not guaranteed since we could have the mirror leptons decay when the SM sphalerons are not in equilibrium and still produce a baryon asymmetry. We leave a more detailed treatment to future work. %Signal regions with displaced vertices have essentially no background~\cite{CMS:2014hka, CMS:2014wda, Aad:2015rba}. So the sensitivity of these searches is comparable to the ones of those for stable charged particles. We leave a more detailed treatment to future work.

%%%%%%%%%%%%%%%%%%%%%%%%%%%%%%%%%%%%%%%%%%
\section{Conclusion}
In this paper, we have shown that solutions to the strong CP problem based on parity are potentially testable at the LHC.  Constraints from higher dimensional operators and the stringent bounds on the neutron EDM force the presence of colored particles with mass smaller than about 10 TeV.  These new particles can be collider stable or decay into the SM quarks through a W, Z or Higgs boson. Current limits on their masses range between $500$~GeV and 1 TeV. The second run of the LHC will explore a larger fraction of parameter space as will future 100 TeV proton colliders~\cite{FCC, SppC}. This works motivates collider searches for weak scale particles based on the strong CP problem, rather than the traditional hierarchy problem or the WIMP miracle. Some of the signatures that we discuss, such as stable heavy quarks and decays to light quarks, are not similar to any of the traditional manifestations of a solution to the hierarchy problem and have never been paramount in the experimental collaborations schedules. 

\acknowledgments
We thank Nima Arkani-Hamed, Loic Quertenmont and Prashant Saraswat for helpful discussions. RTD is supported by the NSF grant PHY-0907744.  AH is supported by the DOE grant DE-SC000998.

\bibliography{ref}

\begin{thebibliography}{47}
\expandafter\ifx\csname natexlab\endcsname\relax\def\natexlab#1{#1}\fi
\expandafter\ifx\csname bibnamefont\endcsname\relax
  \def\bibnamefont#1{#1}\fi
\expandafter\ifx\csname bibfnamefont\endcsname\relax
  \def\bibfnamefont#1{#1}\fi
\expandafter\ifx\csname citenamefont\endcsname\relax
  \def\citenamefont#1{#1}\fi
\expandafter\ifx\csname url\endcsname\relax
  \def\url#1{\texttt{#1}}\fi
\expandafter\ifx\csname urlprefix\endcsname\relax\def\urlprefix{URL }\fi
\providecommand{\bibinfo}[2]{#2}
\providecommand{\eprint}[2][]{\url{#2}}

\bibitem[{\citenamefont{Baker et~al.}(2006)\citenamefont{Baker, Doyle,
  Geltenbort, Green, van~der Grinten et~al.}}]{Baker:2006ts}
\bibinfo{author}{\bibfnamefont{C.}~\bibnamefont{Baker}},
  \bibinfo{author}{\bibfnamefont{D.}~\bibnamefont{Doyle}},
  \bibinfo{author}{\bibfnamefont{P.}~\bibnamefont{Geltenbort}},
  \bibinfo{author}{\bibfnamefont{K.}~\bibnamefont{Green}},
  \bibinfo{author}{\bibfnamefont{M.}~\bibnamefont{van~der Grinten}},
  \bibnamefont{et~al.}, \bibinfo{journal}{Phys.Rev.Lett.}
  \textbf{\bibinfo{volume}{97}}, \bibinfo{pages}{131801}
  (\bibinfo{year}{2006}), \eprint{hep-ex/0602020}.

\bibitem[{\citenamefont{Vicari and Panagopoulos}(2009)}]{Vicari:2008jw}
\bibinfo{author}{\bibfnamefont{E.}~\bibnamefont{Vicari}} \bibnamefont{and}
  \bibinfo{author}{\bibfnamefont{H.}~\bibnamefont{Panagopoulos}},
  \bibinfo{journal}{Phys.Rept.} \textbf{\bibinfo{volume}{470}},
  \bibinfo{pages}{93} (\bibinfo{year}{2009}), \eprint{0803.1593}.

\bibitem[{\citenamefont{Engel et~al.}(2013)\citenamefont{Engel, Ramsey-Musolf,
  and van Kolck}}]{Engel:2013lsa}
\bibinfo{author}{\bibfnamefont{J.}~\bibnamefont{Engel}},
  \bibinfo{author}{\bibfnamefont{M.~J.} \bibnamefont{Ramsey-Musolf}},
  \bibnamefont{and} \bibinfo{author}{\bibfnamefont{U.}~\bibnamefont{van
  Kolck}}, \bibinfo{journal}{Prog.Part.Nucl.Phys.}
  \textbf{\bibinfo{volume}{71}}, \bibinfo{pages}{21} (\bibinfo{year}{2013}),
  \eprint{1303.2371}.

\bibitem[{\citenamefont{Peccei and Quinn}(1977{\natexlab{a}})}]{Peccei:1977hh}
\bibinfo{author}{\bibfnamefont{R.}~\bibnamefont{Peccei}} \bibnamefont{and}
  \bibinfo{author}{\bibfnamefont{H.~R.} \bibnamefont{Quinn}},
  \bibinfo{journal}{Phys.Rev.Lett.} \textbf{\bibinfo{volume}{38}},
  \bibinfo{pages}{1440} (\bibinfo{year}{1977}{\natexlab{a}}).

\bibitem[{\citenamefont{Peccei and Quinn}(1977{\natexlab{b}})}]{Peccei:1977ur}
\bibinfo{author}{\bibfnamefont{R.}~\bibnamefont{Peccei}} \bibnamefont{and}
  \bibinfo{author}{\bibfnamefont{H.~R.} \bibnamefont{Quinn}},
  \bibinfo{journal}{Phys.Rev.} \textbf{\bibinfo{volume}{D16}},
  \bibinfo{pages}{1791} (\bibinfo{year}{1977}{\natexlab{b}}).

\bibitem[{\citenamefont{Weinberg}(1978)}]{Weinberg:1977ma}
\bibinfo{author}{\bibfnamefont{S.}~\bibnamefont{Weinberg}},
  \bibinfo{journal}{Phys.Rev.Lett.} \textbf{\bibinfo{volume}{40}},
  \bibinfo{pages}{223} (\bibinfo{year}{1978}).

\bibitem[{\citenamefont{Wilczek}(1978)}]{Wilczek:1977pj}
\bibinfo{author}{\bibfnamefont{F.}~\bibnamefont{Wilczek}},
  \bibinfo{journal}{Phys.Rev.Lett.} \textbf{\bibinfo{volume}{40}},
  \bibinfo{pages}{279} (\bibinfo{year}{1978}).

\bibitem[{\citenamefont{'t~Hooft}(1976)}]{'tHooft:1976up}
\bibinfo{author}{\bibfnamefont{G.}~\bibnamefont{'t~Hooft}},
  \bibinfo{journal}{Phys.Rev.Lett.} \textbf{\bibinfo{volume}{37}},
  \bibinfo{pages}{8} (\bibinfo{year}{1976}).

\bibitem[{\citenamefont{Nelson}(1984)}]{Nelson:1983zb}
\bibinfo{author}{\bibfnamefont{A.~E.} \bibnamefont{Nelson}},
  \bibinfo{journal}{Phys.Lett.} \textbf{\bibinfo{volume}{B136}},
  \bibinfo{pages}{387} (\bibinfo{year}{1984}).

\bibitem[{\citenamefont{Barr}(1984)}]{Barr:1984qx}
\bibinfo{author}{\bibfnamefont{S.~M.} \bibnamefont{Barr}},
  \bibinfo{journal}{Phys.Rev.Lett.} \textbf{\bibinfo{volume}{53}},
  \bibinfo{pages}{329} (\bibinfo{year}{1984}).

\bibitem[{\citenamefont{Babu and Mohapatra}(1990)}]{Babu:1989rb}
\bibinfo{author}{\bibfnamefont{K.}~\bibnamefont{Babu}} \bibnamefont{and}
  \bibinfo{author}{\bibfnamefont{R.~N.} \bibnamefont{Mohapatra}},
  \bibinfo{journal}{Phys.Rev.} \textbf{\bibinfo{volume}{D41}},
  \bibinfo{pages}{1286} (\bibinfo{year}{1990}).

\bibitem[{\citenamefont{Vecchi}(2014)}]{Vecchi:2014hpa}
\bibinfo{author}{\bibfnamefont{L.}~\bibnamefont{Vecchi}}
  (\bibinfo{year}{2014}), \eprint{1412.3805}.

\bibitem[{\citenamefont{Barr et~al.}(1991)\citenamefont{Barr, Chang, and
  Senjanovic}}]{Barr:1991qx}
\bibinfo{author}{\bibfnamefont{S.~M.} \bibnamefont{Barr}},
  \bibinfo{author}{\bibfnamefont{D.}~\bibnamefont{Chang}}, \bibnamefont{and}
  \bibinfo{author}{\bibfnamefont{G.}~\bibnamefont{Senjanovic}},
  \bibinfo{journal}{Phys.Rev.Lett.} \textbf{\bibinfo{volume}{67}},
  \bibinfo{pages}{2765} (\bibinfo{year}{1991}).

\bibitem[{\citenamefont{Hook}(2014)}]{Hook:2014cda}
\bibinfo{author}{\bibfnamefont{A.}~\bibnamefont{Hook}} (\bibinfo{year}{2014}),
  \eprint{1411.3325}.

\bibitem[{\citenamefont{Ellis and Gaillard}(1979)}]{Ellis:1978hq}
\bibinfo{author}{\bibfnamefont{J.~R.} \bibnamefont{Ellis}} \bibnamefont{and}
  \bibinfo{author}{\bibfnamefont{M.~K.} \bibnamefont{Gaillard}},
  \bibinfo{journal}{Nucl.Phys.} \textbf{\bibinfo{volume}{B150}},
  \bibinfo{pages}{141} (\bibinfo{year}{1979}).

\bibitem[{\citenamefont{Arkani-Hamed and
  Dimopoulos}(2005)}]{ArkaniHamed:2004fb}
\bibinfo{author}{\bibfnamefont{N.}~\bibnamefont{Arkani-Hamed}}
  \bibnamefont{and}
  \bibinfo{author}{\bibfnamefont{S.}~\bibnamefont{Dimopoulos}},
  \bibinfo{journal}{JHEP} \textbf{\bibinfo{volume}{0506}}, \bibinfo{pages}{073}
  (\bibinfo{year}{2005}), \eprint{hep-th/0405159}.

\bibitem[{\citenamefont{Kamionkowski and
  March-Russell}(1992)}]{Kamionkowski:1992mf}
\bibinfo{author}{\bibfnamefont{M.}~\bibnamefont{Kamionkowski}}
  \bibnamefont{and}
  \bibinfo{author}{\bibfnamefont{J.}~\bibnamefont{March-Russell}},
  \bibinfo{journal}{Phys.Lett.} \textbf{\bibinfo{volume}{B282}},
  \bibinfo{pages}{137} (\bibinfo{year}{1992}), \eprint{hep-th/9202003}.

\bibitem[{\citenamefont{Kallosh et~al.}(1995)\citenamefont{Kallosh, Linde,
  Linde, and Susskind}}]{Kallosh:1995hi}
\bibinfo{author}{\bibfnamefont{R.}~\bibnamefont{Kallosh}},
  \bibinfo{author}{\bibfnamefont{A.~D.} \bibnamefont{Linde}},
  \bibinfo{author}{\bibfnamefont{D.~A.} \bibnamefont{Linde}}, \bibnamefont{and}
  \bibinfo{author}{\bibfnamefont{L.}~\bibnamefont{Susskind}},
  \bibinfo{journal}{Phys.Rev.} \textbf{\bibinfo{volume}{D52}},
  \bibinfo{pages}{912} (\bibinfo{year}{1995}), \eprint{hep-th/9502069}.

\bibitem[{\citenamefont{Holman et~al.}(1992)\citenamefont{Holman, Hsu, Kephart,
  Kolb, Watkins et~al.}}]{Holman:1992us}
\bibinfo{author}{\bibfnamefont{R.}~\bibnamefont{Holman}},
  \bibinfo{author}{\bibfnamefont{S.~D.} \bibnamefont{Hsu}},
  \bibinfo{author}{\bibfnamefont{T.~W.} \bibnamefont{Kephart}},
  \bibinfo{author}{\bibfnamefont{E.~W.} \bibnamefont{Kolb}},
  \bibinfo{author}{\bibfnamefont{R.}~\bibnamefont{Watkins}},
  \bibnamefont{et~al.}, \bibinfo{journal}{Phys.Lett.}
  \textbf{\bibinfo{volume}{B282}}, \bibinfo{pages}{132} (\bibinfo{year}{1992}),
  \eprint{hep-ph/9203206}.

\bibitem[{\citenamefont{Berezhiani et~al.}(1993)\citenamefont{Berezhiani,
  Mohapatra, and Senjanovic}}]{Berezhiani:1992pq}
\bibinfo{author}{\bibfnamefont{Z.~G.} \bibnamefont{Berezhiani}},
  \bibinfo{author}{\bibfnamefont{R.~N.} \bibnamefont{Mohapatra}},
  \bibnamefont{and}
  \bibinfo{author}{\bibfnamefont{G.}~\bibnamefont{Senjanovic}},
  \bibinfo{journal}{Phys.Rev.} \textbf{\bibinfo{volume}{D47}},
  \bibinfo{pages}{5565} (\bibinfo{year}{1993}), \eprint{hep-ph/9212318}.

\bibitem[{\citenamefont{Blankenburg et~al.}(2012)\citenamefont{Blankenburg,
  Ellis, and Isidori}}]{Blankenburg:2012ex}
\bibinfo{author}{\bibfnamefont{G.}~\bibnamefont{Blankenburg}},
  \bibinfo{author}{\bibfnamefont{J.}~\bibnamefont{Ellis}}, \bibnamefont{and}
  \bibinfo{author}{\bibfnamefont{G.}~\bibnamefont{Isidori}},
  \bibinfo{journal}{Phys.Lett.} \textbf{\bibinfo{volume}{B712}},
  \bibinfo{pages}{386} (\bibinfo{year}{2012}), \eprint{1202.5704}.

\bibitem[{\citenamefont{Aad et~al.}(2015{\natexlab{a}})}]{ATLAS:2014fka}
\bibinfo{author}{\bibfnamefont{G.}~\bibnamefont{Aad}} \bibnamefont{et~al.}
  (\bibinfo{collaboration}{ATLAS}), \bibinfo{journal}{JHEP}
  \textbf{\bibinfo{volume}{1501}}, \bibinfo{pages}{068}
  (\bibinfo{year}{2015}{\natexlab{a}}), \eprint{1411.6795}.

\bibitem[{\citenamefont{Chatrchyan
  et~al.}(2014{\natexlab{a}})}]{Chatrchyan:2013uxa}
\bibinfo{author}{\bibfnamefont{S.}~\bibnamefont{Chatrchyan}}
  \bibnamefont{et~al.} (\bibinfo{collaboration}{CMS Collaboration}),
  \bibinfo{journal}{Phys.Lett.} \textbf{\bibinfo{volume}{B729}},
  \bibinfo{pages}{149} (\bibinfo{year}{2014}{\natexlab{a}}),
  \eprint{1311.7667}.

\bibitem[{\citenamefont{Aliev et~al.}(2011)\citenamefont{Aliev, Lacker,
  Langenfeld, Moch, Uwer et~al.}}]{Aliev:2010zk}
\bibinfo{author}{\bibfnamefont{M.}~\bibnamefont{Aliev}},
  \bibinfo{author}{\bibfnamefont{H.}~\bibnamefont{Lacker}},
  \bibinfo{author}{\bibfnamefont{U.}~\bibnamefont{Langenfeld}},
  \bibinfo{author}{\bibfnamefont{S.}~\bibnamefont{Moch}},
  \bibinfo{author}{\bibfnamefont{P.}~\bibnamefont{Uwer}}, \bibnamefont{et~al.},
  \bibinfo{journal}{Comput.Phys.Commun.} \textbf{\bibinfo{volume}{182}},
  \bibinfo{pages}{1034} (\bibinfo{year}{2011}), \eprint{1007.1327}.

\bibitem[{\citenamefont{Chatrchyan
  et~al.}(2013{\natexlab{a}})}]{Chatrchyan:2013oca}
\bibinfo{author}{\bibfnamefont{S.}~\bibnamefont{Chatrchyan}}
  \bibnamefont{et~al.} (\bibinfo{collaboration}{CMS}), \bibinfo{journal}{JHEP}
  \textbf{\bibinfo{volume}{1307}}, \bibinfo{pages}{122}
  (\bibinfo{year}{2013}{\natexlab{a}}), \eprint{1305.0491}.

\bibitem[{\citenamefont{Mackeprang and Milstead}(2010)}]{Mackeprang:2009ad}
\bibinfo{author}{\bibfnamefont{R.}~\bibnamefont{Mackeprang}} \bibnamefont{and}
  \bibinfo{author}{\bibfnamefont{D.}~\bibnamefont{Milstead}},
  \bibinfo{journal}{Eur.Phys.J.} \textbf{\bibinfo{volume}{C66}},
  \bibinfo{pages}{493} (\bibinfo{year}{2010}), \eprint{0908.1868}.

\bibitem[{\citenamefont{Kraan}(2004)}]{Kraan:2004tz}
\bibinfo{author}{\bibfnamefont{A.~C.} \bibnamefont{Kraan}},
  \bibinfo{journal}{Eur.Phys.J.} \textbf{\bibinfo{volume}{C37}},
  \bibinfo{pages}{91} (\bibinfo{year}{2004}), \eprint{hep-ex/0404001}.

\bibitem[{\citenamefont{Mackeprang and Rizzi}(2007)}]{Mackeprang:2006gx}
\bibinfo{author}{\bibfnamefont{R.}~\bibnamefont{Mackeprang}} \bibnamefont{and}
  \bibinfo{author}{\bibfnamefont{A.}~\bibnamefont{Rizzi}},
  \bibinfo{journal}{Eur.Phys.J.} \textbf{\bibinfo{volume}{C50}},
  \bibinfo{pages}{353} (\bibinfo{year}{2007}), \eprint{hep-ph/0612161}.

\bibitem[{\citenamefont{Aad et~al.}(2015{\natexlab{b}})}]{ATLAS-CONF-2015-012}
\bibinfo{author}{\bibfnamefont{G.}~\bibnamefont{Aad}} \bibnamefont{et~al.}
  (\bibinfo{collaboration}{ATLAS}) (\bibinfo{year}{2015}{\natexlab{b}}),
  \eprint{ATLAS-CONF-2015-012, ATLAS-COM-CONF-2015-012}.

\bibitem[{\citenamefont{Chatrchyan et~al.}(2012{\natexlab{a}})}]{CMS:yut}
\bibinfo{author}{\bibfnamefont{S.}~\bibnamefont{Chatrchyan}}
  \bibnamefont{et~al.} (\bibinfo{collaboration}{CMS})
  (\bibinfo{year}{2012}{\natexlab{a}}), \eprint{CMS-PAS-SUS-12-027}.

\bibitem[{\citenamefont{Chatrchyan et~al.}(2014{\natexlab{b}})}]{CMS:2014dka}
\bibinfo{author}{\bibfnamefont{S.}~\bibnamefont{Chatrchyan}}
  \bibnamefont{et~al.} (\bibinfo{collaboration}{CMS})
  (\bibinfo{year}{2014}{\natexlab{b}}), \eprint{CMS-PAS-B2G-12-017}.

\bibitem[{\citenamefont{Van~Onsem}(2014)}]{VanOnsem:2014dba}
\bibinfo{author}{\bibfnamefont{G.~P.} \bibnamefont{Van~Onsem}}
  (\bibinfo{year}{2014}), \eprint{CERN-THESIS-2014-108, CMS-TS-2014-024}.

\bibitem[{\citenamefont{Aad et~al.}(2012{\natexlab{a}})}]{ATLAS:2012apa}
\bibinfo{author}{\bibfnamefont{G.}~\bibnamefont{Aad}} \bibnamefont{et~al.}
  (\bibinfo{collaboration}{ATLAS}) (\bibinfo{year}{2012}{\natexlab{a}}),
  \eprint{ATLAS-CONF-2012-137, ATLAS-COM-CONF-2012-167}.

\bibitem[{\citenamefont{Aad et~al.}(2012{\natexlab{b}})}]{Aad:2011yn}
\bibinfo{author}{\bibfnamefont{G.}~\bibnamefont{Aad}} \bibnamefont{et~al.}
  (\bibinfo{collaboration}{ATLAS}), \bibinfo{journal}{Phys.Lett.}
  \textbf{\bibinfo{volume}{B712}}, \bibinfo{pages}{22}
  (\bibinfo{year}{2012}{\natexlab{b}}), \eprint{1112.5755}.

\bibitem[{\citenamefont{Chatrchyan et~al.}(2013{\natexlab{b}})}]{CMS:2013fea}
\bibinfo{author}{\bibfnamefont{S.}~\bibnamefont{Chatrchyan}}
  \bibnamefont{et~al.} (\bibinfo{collaboration}{CMS})
  (\bibinfo{year}{2013}{\natexlab{b}}), \eprint{CMS-PAS-EXO-12-024}.

\bibitem[{\citenamefont{Aad et~al.}(2013)}]{TheATLAScollaboration:2013xia}
\bibinfo{author}{\bibfnamefont{G.}~\bibnamefont{Aad}} \bibnamefont{et~al.}
  (\bibinfo{collaboration}{ATLAS}) (\bibinfo{year}{2013}),
  \eprint{ATLAS-CONF-2013-091, ATLAS-COM-CONF-2013-081}.

\bibitem[{\citenamefont{Chatrchyan
  et~al.}(2014{\natexlab{c}})}]{Chatrchyan:2014aea}
\bibinfo{author}{\bibfnamefont{S.}~\bibnamefont{Chatrchyan}}
  \bibnamefont{et~al.} (\bibinfo{collaboration}{CMS}),
  \bibinfo{journal}{Phys.Rev.} \textbf{\bibinfo{volume}{D90}},
  \bibinfo{pages}{032006} (\bibinfo{year}{2014}{\natexlab{c}}),
  \eprint{1404.5801}.

\bibitem[{\citenamefont{Chatrchyan et~al.}(2012{\natexlab{b}})}]{CMS:zva}
\bibinfo{author}{\bibfnamefont{S.}~\bibnamefont{Chatrchyan}}
  \bibnamefont{et~al.} (\bibinfo{collaboration}{CMS})
  (\bibinfo{year}{2012}{\natexlab{b}}), \eprint{CMS-PAS-EXO-12-042}.

\bibitem[{\citenamefont{Blum et~al.}(2015)\citenamefont{Blum, D’Agnolo, and
  Fan}}]{Blum:2015rpa}
\bibinfo{author}{\bibfnamefont{K.}~\bibnamefont{Blum}},
  \bibinfo{author}{\bibfnamefont{R.~T.} \bibnamefont{D’Agnolo}},
  \bibnamefont{and} \bibinfo{author}{\bibfnamefont{J.}~\bibnamefont{Fan}},
  \bibinfo{journal}{JHEP} \textbf{\bibinfo{volume}{1503}}, \bibinfo{pages}{166}
  (\bibinfo{year}{2015}), \eprint{1502.01045}.

\bibitem[{\citenamefont{Delaunay et~al.}(2014)\citenamefont{Delaunay, Flacke,
  Gonzalez-Fraile, Lee, Panico et~al.}}]{Delaunay:2013pwa}
\bibinfo{author}{\bibfnamefont{C.}~\bibnamefont{Delaunay}},
  \bibinfo{author}{\bibfnamefont{T.}~\bibnamefont{Flacke}},
  \bibinfo{author}{\bibfnamefont{J.}~\bibnamefont{Gonzalez-Fraile}},
  \bibinfo{author}{\bibfnamefont{S.~J.} \bibnamefont{Lee}},
  \bibinfo{author}{\bibfnamefont{G.}~\bibnamefont{Panico}},
  \bibnamefont{et~al.}, \bibinfo{journal}{JHEP}
  \textbf{\bibinfo{volume}{1402}}, \bibinfo{pages}{055} (\bibinfo{year}{2014}),
  \eprint{1311.2072}.

\bibitem[{\citenamefont{Cooke et~al.}(2013)\citenamefont{Cooke, Pettini,
  Jorgenson, Murphy, and Steidel}}]{Cooke:2013cba}
\bibinfo{author}{\bibfnamefont{R.}~\bibnamefont{Cooke}},
  \bibinfo{author}{\bibfnamefont{M.}~\bibnamefont{Pettini}},
  \bibinfo{author}{\bibfnamefont{R.~A.} \bibnamefont{Jorgenson}},
  \bibinfo{author}{\bibfnamefont{M.~T.} \bibnamefont{Murphy}},
  \bibnamefont{and} \bibinfo{author}{\bibfnamefont{C.~C.}
  \bibnamefont{Steidel}} (\bibinfo{year}{2013}), \eprint{1308.3240}.

\bibitem[{\citenamefont{Ade et~al.}(2015)}]{Ade:2015xua}
\bibinfo{author}{\bibfnamefont{P.}~\bibnamefont{Ade}} \bibnamefont{et~al.}
  (\bibinfo{collaboration}{Planck}) (\bibinfo{year}{2015}),
  \eprint{1502.01589}.

\bibitem[{\citenamefont{Khachatryan et~al.}(2015{\natexlab{a}})}]{CMS:2014hka}
\bibinfo{author}{\bibfnamefont{V.}~\bibnamefont{Khachatryan}}
  \bibnamefont{et~al.} (\bibinfo{collaboration}{CMS}),
  \bibinfo{journal}{Phys.Rev.} \textbf{\bibinfo{volume}{D91}},
  \bibinfo{pages}{052012} (\bibinfo{year}{2015}{\natexlab{a}}),
  \eprint{1411.6977}.

\bibitem[{\citenamefont{Khachatryan et~al.}(2015{\natexlab{b}})}]{CMS:2014wda}
\bibinfo{author}{\bibfnamefont{V.}~\bibnamefont{Khachatryan}}
  \bibnamefont{et~al.} (\bibinfo{collaboration}{CMS}),
  \bibinfo{journal}{Phys.Rev.} \textbf{\bibinfo{volume}{D91}},
  \bibinfo{pages}{012007} (\bibinfo{year}{2015}{\natexlab{b}}),
  \eprint{1411.6530}.

\bibitem[{\citenamefont{Aad et~al.}(2015{\natexlab{c}})}]{Aad:2015rba}
\bibinfo{author}{\bibfnamefont{G.}~\bibnamefont{Aad}} \bibnamefont{et~al.}
  (\bibinfo{collaboration}{ATLAS}) (\bibinfo{year}{2015}{\natexlab{c}}),
  \eprint{1504.05162}.

\bibitem[{FCC()}]{FCC}
\bibinfo{note}{Future Circular Collider Study Kick-Off meeting},
  \urlprefix\url{https://indico.cern.ch/event/282344/}.

\bibitem[{Spp()}]{SppC}
\bibinfo{note}{Workshop on Future High Energy Circular Colliders},
  \urlprefix\url{http://indico.ihep.ac.cn/conferenceDisplay.py?confId=3813}.

\end{thebibliography}

\end{document}